\documentclass[aps,prl,twocolumn,superscriptaddress,showpacs]{revtex4}
\usepackage{mathrsfs}
\usepackage{graphicx}  % Include figure files
\usepackage{cases}

\begin{document}

\title{Thermalization Breakdown and Conductivity Improvement within the Interacting Dynamic Disorder Model}

\author{Yao Yao}
\email[Electronic mail: ] {yaoyao@fudan.edu.cn} \affiliation{State
Key Laboratory of Surface Physics and Department of Physics, Fudan
University, Shanghai 200433, China}
\author{Yiqiang Zhan}
\affiliation{State Key Laboratory of ASIC and System, Department of
Microelectronics, SIST, Fudan University, Shanghai 200433, China}
\author{Xiaoyuan Hou}
\affiliation{State Key Laboratory of Surface Physics and Department
of Physics, Fudan University, Shanghai 200433, China}
\author{Chang-Qin Wu}
\affiliation{State Key Laboratory of Surface Physics and Department
of Physics, Fudan University, Shanghai 200433, China}

\date{\today}
\begin{abstract}
Based on the framework of Kubo formulism, we develop the minimally
entangled typical thermal state algorithm to study the temperature
and time dependence of current-current correlation function in
one-dimensional spinless fermion model, taking into account both the
electron-electron (e-e) intersite interaction and the dynamic
disorder induced by classical phonons. Without e-e interaction, the
numerical results, showing an exponential decay of the time
dependent correlation, could be precisely compared with that from
the analytical derivation, namely, from the generalized Langevin
equation. More importantly, when a strong enough e-e interaction is
presence, we find a long-time correlation in the regime of small
dynamic disorder, indicating the breakdown of thermal relaxation,
which is a typical many-body effect. On the basis of this finding,
we show that it might be applied to understand the metalliclike
charge transport and the abnormal improvement of the conductivity
with respect to the redoping experiment in K$_3$C$_{60}$, an organic
superconducting material.
\end{abstract}


\maketitle

Recent progresses of organic superconducting materials, such as
potassium-doped picene,\cite{SC1,SC2} phenanthrene,\cite{SC3}
coronene,\cite{SC4} and dibenzpentacene,\cite{SC5} has opened a new
research subfield in organic electronics, due to the following two
critical points: The electron-intramolecular-vibration
interaction,\cite{SCT4,SCT7} together with the intercalant and
intermolecular phonons,\cite{SCT1,SCT5,SCT6} are substantially
respondence to the superconductivity; The materials are typical
strongly-correlated electron systems.\cite{SCT6,SCT2,SCT3} Both the
two statements are from first-principle calculations, and an
in-depth model computation is not found. On the other hand, even
under room temperature, once doped with alkali metal, those
originally semiconducting materials become to behave metalliclike
conductivity.\cite{SC1,K3C601,K3C602} Intuitively, the doping of
alkali metal has modified the $\pi$-electron structure of the
organic molecules,\cite{SC1} and the electron-electron (e-e) Coulomb
interaction becomes to improve the conductivity. But in a classical
manner, the e-e interaction always plays a negative role (blocking)
in the charge transport.\cite{cou} This contradiction implies that,
a completely quantum description should be addressed for this
subject, which is the main motivation of this work.

Beyond (semi-)classical treatment, there have been many quantum
theories for the transport in organic
solids.\cite{theory1,theory2,theory3,theory4,troisi,troisi2,SF1,SF2,theory5}
Most of the works paid attention to the subjects, such as the
unified bandlike and hopping
mobility,\cite{theory1,theory2,theory3,theory4} the static and
dynamic disorder,\cite{troisi,troisi2,SF1,SF2} and the mixed quantum
and classical problems.\cite{theory5} In particular, the dynamic
disorder model,\cite{troisi} based upon the one-dimensional
Holstein-Peierls Hamiltonian with both intra- and inter-molecular
phonons treated classically, was extensively used to comprehensively
understand the behavior of transport in organic crystals.
Originally, one used the Ehrenfest method to simulate the diffusion
behavior of an initially localized electron
wavepacket\cite{troisi,troisi2,SF1} and successfully given the basic
carrier's bandlike mobility.\cite{troisi2} However, as those works
were mainly working within one-particle picture, they make no sense
of the fluctuation of the particle number, so that the dynamical
response could not be evaluated. To fix this problem, the Kubo
formulism was taken into account, in which the mobility is directly
related to the current-current correlation
function.\cite{SF2,theory2} It was obtained that, both bandlike and
hopping transport could be described, that is, the localization
length decreased quickly when the electron-phonon (e-p) coupling
increases.\cite{SF2} At the mean time, the e-e interaction is also
studied on the mean field level.\cite{theory4} In addition, the 1D
Holstein model was also applied to study the organic
superconductivity, since the e-p coupling is recognized to be mostly
relevant.\cite{bursill} In all, it seems to say that, one can just
straightforwardly follow this line to study the e-e correlation more
comprehensively, which should be the essential character of organic
superconducting materials. Yet, as we will show in this Letter, the
breakdown of thermalization\cite{TB1,TB2,TB3} induced by the e-e
correlation makes the problem quite novel.

The density matrix renormalization group (DMRG), a well-known
numerical method, is one of the most powerful methods to deal with
the one-dimensional strongly-correlated systems.\cite{Review} In the
last decade, lots of effort have been put into extending the method
to finite-temperature
problems.\cite{TDMRG1,TDMRG2,TDMRG3,TDMRG4,METTS} Advantages from
White, who is the inventor of DMRG, are made by introducing the
language of matrix product state and quantum Monte Carlo method, and
a so-called minimally entangled typical thermal state (METTS)
algorithm was established.\cite{METTS} This new method is highly
efficient to calculate the thermal quantities of the system of
one-dimensional spin (and thus spinless fermion) lattices,
especially under high temperature. Furthermore, if the
imaginary-time evolution operators in METTS algorithm are replaced
by its real-time counterpart, which is quite straightforward, the
method is then applicable for both temperature and time dependent
problems. This means it finally becomes possible to study the
thermodynamics of an organic electronic systems, such as Kubo
formula and time-dependent current-current correlation, with both
e-p and e-e interaction presence.

The model we are dealing with is a one-dimensional spinless model
with near-neighboring e-e interactions, and the Holstein e-p
coupling, which is treated classically, is also taken into account
to act as a dynamic disorder. The Hamiltonian writes,
\begin{eqnarray}
H=H_{\rm el}+H_{\rm ph}.\label{hami}
\end{eqnarray}
The electronic part is
\begin{eqnarray}
H_{\rm el}&=&-t_0\sum_{j}(c^\dagger_{j+1}c_{j}+{\rm
h.c.})+g\sum_{j}u_j\hat{n}_j\nonumber\\&+&V\sum_j\hat{n}_j\hat{n}_{j+1},
\end{eqnarray}
where $c^\dagger_{j}(c_{j})$ creates (annihilates) an electron on
the $j$-th site, $u_j$ the displacement of the $j$-th site,
$\hat{n}_j(\equiv c^\dagger_{j}c_{j}$) the number operator of the
electron, $t_0$ the transfer integral, $g$ the e-p coupling
constant, and $V$ the intersite e-e interaction. The phonon part of
Hamiltonian (\ref{hami}) is described as
\begin{eqnarray}
H_{\rm
ph}=\frac{K}{2}\sum_{j}u_{j}^2+\frac{M}{2}\sum_{j}\dot{u}_j^2,
\end{eqnarray}
where $K$ is the elastic constant between neighbor sites, and $M$
the mass of a site. All the parameters in the model could be
determined by first-principle calculations. For example, the
intermolecular transfer integral $t_0$ in doped corenene is around
$30$meV,\cite{SCT4} and the characteristic phonon frequency in doped
picene is around $18$meV.\cite{SCT1} But in this work, we will take
the dimensionless parameters, that is, $t, K$ and $M$ are all set to
unity, such that the frequency of phonons is $\omega=\sqrt{K/M}=1$.
$g$ and $V$ will be the main adjustable parameters, and the main
results are calculated in an open chain with $60$ sites. To diminish
the influence of open boundary condition, we have made the $t_0$
exponential decay on several bonds near the boundary. Actually, the
1D spinless fermion model with e-p coupling has been long-termly
studied to understand the superconductivity in doped
fullerenes.\cite{bursill} It was found that, the phase transition
from Luttinger liquid phase to charge density wave occurs around
$g=2t_0$ when $t\geq1$. This critical phenomenon is also found in
our calculations, but the main physical results in this work are
within the Luttinger liquid phase.

%could induce the pronounced many-body correlated effects like in
%Hubbard model [] except the spin-charge separation, which is not
%concerned on the present stage. The temperature changes in a large
%extent from 100 to 300K. Due to the limit of method, we can not
%reach the lower temperature, under which superconductivity happens.

Based on the Hamiltonian, we are going to calculate the
zero-frequency Kubo formula\cite{Kubo} defined as
\begin{eqnarray}
\sigma=\frac{1}{k_BT}\int_{0}^{\infty}dt\langle
\hat{I}(t)\hat{I}(0)\rangle,\label{kubo}
\end{eqnarray}
where $\sigma$ is the conductivity, $T$ the temperature,
$\hat{I}(t)$ ($\equiv e^{iHt/\hbar}\hat{I}e^{-iHt/\hbar}$) the
current operator, and $\langle \hat{A}\rangle$ the thermal average
defined as $\langle \hat{A}\rangle={\rm
Tr}(e^{-H/k_BT}\hat{A})/\mathcal{Z}$ with $\mathcal{Z}$ the
partition function. Herein, to define the current operator
$\hat{I}$, one should first define the polarization operator
$\hat{P}$ of the center of mass of the electrons, namely,
\begin{eqnarray}
\hat{P}=e\sum_j R_jc^{\dag}_jc_j,
\end{eqnarray}
with $e$ the charge of electron and $R_j$ the position of each site.
Then $\hat{I}$ could be defined by
\begin{eqnarray}
\hat{I}=\frac{d\hat{P}}{dt}=\frac{1}{i\hbar}[\hat{P},H]=-\frac{eat_0}{i\hbar}\sum_j
(c^{\dag}_jc_{j+1}-{\rm h.c.}),
\end{eqnarray}
with $a$ the lattice constant.

The current-current correlation function in the Kubo formula
(\ref{kubo}) is both time and temperature dependent. In order to
evaluate it, we apply the METTS algorithm combined with
time-dependent DMRG method (tDMRG),\cite{timeDMRG} whose basic
procedure is as follows. Firstly, one initializes a configuration of
the occupation of each site arbitrarily and then produces a
so-called classical product state (CPS) as
\begin{eqnarray}
|n\rangle=|n_1,n_2,\cdot\cdot\cdot n_j\cdot\cdot\cdot\rangle,
\end{eqnarray}
with $|n_j\rangle$ the local basis on site $j$. Secondly, the
imaginary-time evolution operator is acting on the CPS, namely,
\begin{eqnarray}
|\phi_n\rangle=P(n)^{-1/2}e^{-\beta H/2}|n\rangle,\label{Im}
\end{eqnarray}
where $|\phi_n\rangle$ is the so-called typical thermal state,
$P(n)(\equiv \langle n|e^{-\beta H}|n\rangle)$ the statistical
probability of the state $|n\rangle$ in the ensemble, and $\beta$
the inverse of $k_BT$. In a Monte Carlo manner, the above steps are
iterated, and then we get lots of (but still much fewer than the
total number of state) $|\phi_n\rangle$ samplings in hand. Based on
these samplings, we calculate the expectation of any operator $A$ as
\begin{eqnarray}
\langle A\rangle=\frac{1}{\mathcal{Z}}\sum_n
P(n)\langle\phi_n|A|\phi_n\rangle.
\end{eqnarray}
Obviously, when the number of site becomes large, the Hilbert space
is enlarged drastically. Hence, one should follow the standard
procedure of tDMRG, that is, the evolution operator must be
decomposed onto individual bonds, and the state must be truncated
while scanning as the usual procedure in DMRG method. Meanwhile, in
White's treatment, to ensure the ergodic hypothesis, one should
choose another CPS by collapsing $|\phi_n\rangle$ to (the arbitrary
basis of) each site and iterate the above steps for sufficient
times.

Once the thermal equilibrium is reached, that is, the typical
thermal states $|\phi_n(0)\rangle$ in the equilibrium are obtained,
one can easily calculate the time evolution by changing the
evolution operator with imaginary time in (\ref{Im}) to real time.
For the aim of the present work, we are focusing on computing the
time dependent current-current correlation function in Kubo formula,
i.e.,
$\langle\phi_n(0)|e^{iHt/\hbar}\hat{I}e^{-iHt/\hbar}\hat{I}|\phi_n(0)\rangle$.
Hence, one should first make the current operator $\hat{I}$ acting
onto $|\phi_n(0)\rangle$, and then calculate the time evolution of
the obtained state. At each time step $t$, one again act $\hat{I}$
on the new state and calculate its overlap with the state
$e^{-iHt/\hbar}|\phi_n(0)\rangle$. Since $\hat{I}$ could be
decomposed onto each bond of the lattice, naively, one can just act
it on the individual bond in each scanning step and sum up the
targeting states. But this idea is quite inefficient for tDMRG,
because we need to target too many states. Our treatment is on the
basis of assumption of translating invariance, namely, at the
initial moment, we only act the current operator on the central bond
of $|\phi_n(0)\rangle$, which should be the most precise bond in the
approximation of tDMRG. Based on this simplicity, only two states
$|\phi_n(0)\rangle$ and $\hat{I}_c|\phi_n(0)\rangle$ (with
$\hat{I}_c$ the current operator on the central bond) are necessary
to be targeted, which makes the procedure much more efficient.

Up to now, the remaining thing is to treat the phonon part of the
system. We follow the usual procedure of Ehrenfest method, which is
the standard method for calculating the time evolution of mixed
classical-quantum systems. That is, we first choose an initial
configuration of $\{u_j\}$ and $\{\dot{u}_j\}$ from the Gaussian
distribution with variance $k_BT/K$ and $k_BT/M$, say the
equilibrium distribution of vibrations. This configuration could be
substituted into the Hamiltonian of electron part, and one follows
the common procedure to calculate the Hellman-Feynman force that
electrons act on the sites, such that, the influence of phonon comes
into the theory.

Here, we need to say that, since the current operator $\hat{I}$ does
not change the total number of electron and the spin degree of
freedom is ignored, the precision of the present numerical
calculation should be at least in the same order with the previous
tDMRG works of us in the Hubbard chain and the conjugated
polymer.\cite{Mine2} However, the evolution is now for the typical
thermal states rather than the ground state. Together with the
numerical errors from the Ehrenfest method for the phonon part, the
accumulation of error is very fast. Even if we decrease the number
of truncated states in METTS, it seems no significant improvement is
obtained, so that more efforts should be devoted to the method
itself. But still, within short time scale, many interesting results
have been found.

\begin{figure}
\includegraphics[angle=0,scale=1.2]{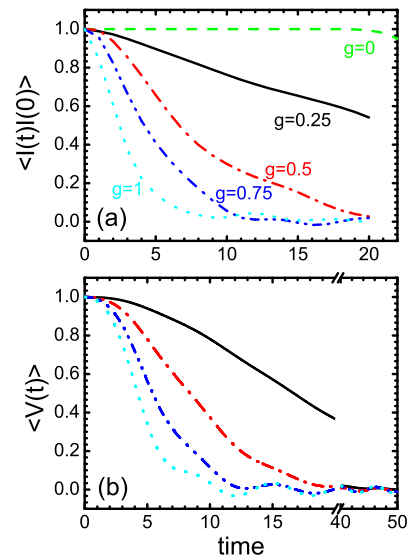}
\caption{(a) The current-current correlation function (in the unit
of $(eat_0)^2/2N_c\hbar^2$) versus time (in the unit of $\hbar/t_0$)
with various e-p coupling. (b) The averaged velocity from the
generalized Langevin equation.\cite{GLE} The relaxation time for
$g=0.25$ is around $50$.}\label{V0}
\end{figure}

In Fig. \ref{V0}(a), we first show the results of current-current
correlation function without e-e interaction under $k_BT=2$. For
$g=0$, i.e., the disorder is absent, the electron is completely
free. One can easily prove that, in this case, $\hat{I}$ commutes
with $H$, so the correlation function is unchanged with time. As we
see that, the curve remains a constant up to $t\doteq20$, but due to
the accumulation of numerical error, the curve goes down quickly
after that point. Hence, $t=20$ should be a point of justification
that, before it the numerical results are credible. In addition, we
find that, following $g$ increasing, the decay of the correlation
becomes much faster, and above $g=2$ (not shown), the correlation
acts nearly a sudden quenching, which means it is within an
insulating phase.

To check the correctness of the present numerical algorithm, a
method beyond the framework of Monte Carlo sampling is needed. Here,
we adopt the generalized Langevin equation (GLE) to calculate the
time dependence of thermal averaged velocity.\cite{GLE} Although not
exactly the same, the oscillation and decay of the averaged velocity
are expected to follow the similar trajectory with that of
current-current correlation function. To compare with the METTS
results, in Fig. \ref{V0}(b), we show the result of $V(t)$ from GLE
with the same $g$'s, respectively. Qualitatively, both the wavy and
decay shape agree with the result from METTS. And more importantly,
when $g$ decreases by the same times, the decay velocity shows very
close value, which could not be adjusted by any parameters. These
results not only allow us to doubly check the correctness of the
numerical algorithm, but also provide more information that METTS
can not give. For example, the GLE result shows the relaxation
behavior for all $g$ after a long time evolution with the relaxation
time is around 50. This complement is very important to understand
the following results.

\begin{figure}
\includegraphics[angle=0,scale=0.9]{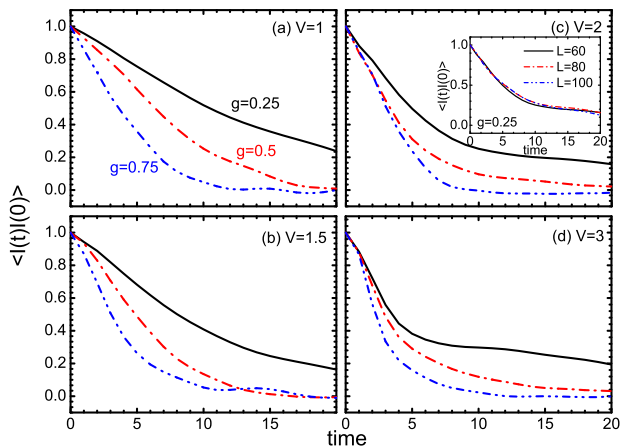}
\caption{Time dependence of current-current correlation function for
various $g$ and $V$. Inset of (c) shows the little influence of the
finite length scale.}\label{ee}
\end{figure}

Now, we are on the stage to show the influence of e-e interaction.
In Fig. \ref{ee}, we show the time dependent correlation for various
$V$ and $g$. Compared with $V=0$ case, when $g$ is larger than
$0.5$, the velocity of decay increases with increasing $V$. This is
easy to understand, since the system tend to the insulating phase.
But quite interestingly, as one can see from (a) to (d) that, when
$g=0.25$, although the velocity of decay still tends to increase in
the very beginning of evolution, it seems that the curve does not
vanish but relax to a finite value (around $0.15$) after a long
time. Of course, one may argue that, our numerical method can not
show the result of long-time evolution, but we can still estimate
from the present result that, the relaxation time for $V=2$ should
be at least much longer than that in uncorrelated system ($\gg50$).
Meanwhile, to exclude the influence of the chain length, in the
inset of Fig. \ref{ee}(c), we show the result for different site
number, and obviously, they are almost the same. Hence, this
long-time memory effect of correlation function is the main finding
of this work. It is non-asymptotic from the uncorrelated system,
since it implies the thermalization breakdown induced by the e-e
interaction.\cite{TB1} Actually, the breakdown of thermal relaxation
in fermion systems was widely studied very recently.\cite{TB2,TB3}
That is, when the main-body correlation enters into the
low-dimensional fermion system, the relaxation to thermal
equilibrium should be strongly related to the initial state, due to
the complex entanglement between equilibrium and nonequilibrium
correlations in correlated systems.\cite{TB2} The present results
agree with these works and add more information on this subject.

\begin{figure}
\includegraphics[angle=0,scale=0.7]{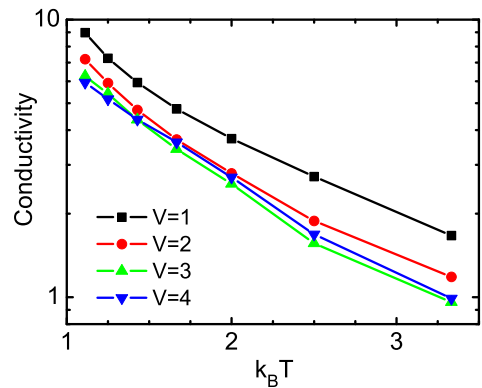}
\caption{Temperature dependence of conductivity (in the unit of
$t_0(ea)^2/2\hbar$) calculated from Kubo formula for various
$V$.}\label{Tdepen}
\end{figure}

Intuitively, since the conductivity is related to the integral of
current-current correlation function within the framework of Kubo
formula, the long-time memory effect found above should contribute a
lot to the conductivity and improve it. This statement obviously
contradicts to the conventional point of view, as the Coulomb
repulsion in the material with small disorder is always recognized
to act as a blocking and thus will lead to the reduction of
conductivity.\cite{cou} Whereas, the improvement of conductivity was
indeed found in K$_3$C$_{60}$ by redoping of alkali metals, as shown
in \cite{K3C601}. The redoping process is, in our opinion, to
increase the concentration of alkali metallic atoms and thus the
Coulomb interaction among them. The improvement of conductivity is
surprising and full of meaning, since it is closely related to the
increase of Tc of these organic superconducting
materials.\cite{K3C601} On the basis of our present finding of the
memory effect, we could then provide a possible explanation on the
experiment. In Fig. \ref{Tdepen}, we show the main results of the
temperature dependence of conductivity. Here, due to the limit of
numerical method, the time integral in the Kubo formula (\ref{kubo})
is only within $t<20$. This is equivalent to considering an
additional static disorder $1/\tau_s$, i.e., the time integral in
Kubo formula becomes\cite{theory2}
\begin{eqnarray}
\int dt\rightarrow\int dte^{-(t/\tau_s)^2}.
\end{eqnarray}
It is found that, when $V<2$, the result is quite the same with the
bandlike behavior in common dynamic disorder model.\cite{troisi}
When $V\geq2$, the relationship between conductivity and temperature
becomes to metalliclike, namely, $\sigma\sim 1/T$. More importantly,
the curve of conductivity stops decreasing but becomes close with
each other. Even in the high temperature regime and $V=4$, we find a
cross of the curves and improvement of the conductivity, which are
consistent with that of doped and annealed cases as shown in the
Fig. 3 of \cite{K3C601}. Therefore, we could now conclude that, the
thermalization breakdown induced by the e-e interaction matters in
the charge transport of alkali-metal-doped organic materials.

Finally, we would like to briefly discuss the superconductivity in
organic solids. Different from their inorganic counterparts, organic
materials have much more diverse vibrational modes, such that the
decoherence process could easily kill the phase correlation of the
electron wavefunctions and thus the tendency of superconductivity.
To avoid this effect, the high frequency part of the phonons, which
is the main source of decoherence,\cite{Mine,Anderson} must be
largely suppressed. The present work provides a possibility, that
is, the long-time memory of electric current induced by e-e
interaction should be a shield against the decoherence and
contribution to the conduction of these materials.

In summary, we have used the METTS algorithm to calculate the
temperature and time dependent current-current correlation function
in a one-dimensional spinless model with both e-e and e-p
interaction taken into account. Via the comparison with analytic
results, we state that, this very new method works well within a
short time scale. Then we study the influence of e-e interaction and
find a long-time memory effect, say, the thermalization breakdown of
the system. Based on this finding, we calculate the temperature
dependent mobility and show that, following the increase of e-e
interaction, the mobility will behave a slight enhancement, which
was also found in the experiment.

\begin{acknowledgments}
This work was supported by the NSF of China, the National Basic
Research Program of China (2009CB29204 and 2012CB921400), and the EC
Project OFSPIN (NMP3-CT-2006-033370).
\end{acknowledgments}

\end{document}

\title{Thermalization Breakdown and Conductivity Improvement within the Interacting Dynamic Disorder Model: Supplementary Material}

\author{Yao Yao}
\email[Electronic mail: ] {yaoyao@fudan.edu.cn} \affiliation{State
Key Laboratory of Surface Physics and Department of Physics, Fudan
University, Shanghai 200433, China}
\author{Xiaoyuan Hou}
\affiliation{State Key Laboratory of Surface Physics and Department
of Physics, Fudan University, Shanghai 200433, China}
\author{Chang-Qin Wu}
\affiliation{State Key Laboratory of Surface Physics and Department
of Physics, Fudan University, Shanghai 200433, China}

\date{\today}

\maketitle

In order to make sense of the long-time behavior for $V=0$, we adopt
the generalized Langevin equation (GLE).\cite{GLE,JCP} The basic
idea is to separate the electronic Hamiltonian into the
center-of-mass and relative part, namely,
\begin{eqnarray}
H=\frac{{
\hat{P}^2}}{2M_e}+\sum_{k}\epsilon_k\tilde{c}^\dagger_{k}\tilde{c}_{k}+g\sum_{q}e^{iq{\rm
\hat{R}}}u_q\tilde{\rho}_q,\label{hami}
\end{eqnarray}
where ${\hat{P}}$ is the momentum of the center of mass, ${\hat{R}}$
the conjugate variable of ${\hat{P}}$, $M_e=Nm_e$ with $N$ the
number of electron and $m_e$ the mass of a single electron,
$\epsilon_k$ the energy with $k$ the wave vector of "relative"
electrons, $\tilde{c}^\dagger_{k}(\tilde{c}_{k})$ the creation
(annihilation) operator of relative electrons with $\tilde{\rho}_q
(\equiv\sum_k\tilde{c}^\dagger_{k+q}\tilde{c}_{k})$ the
corresponding density operator, and $q$ the wave number. It could be
demonstrated that, the operators of center-of-mass and relative part
commute with each other.\cite{GLE2} Meanwhile, only the contribution
of small $q$ is assumed to be dominant, so that the relative part
acts as a source of fluctuations for the center of mass via the
assistance of phonons.

Within this Hamiltonian, we are going to evaluate the derivative of
$\hat{V} (\equiv{\hat{P}}/M_e)$ at time $t$. In the interaction
picture, it could be written as ($\hbar=1$)\cite{GLE}
\begin{eqnarray}
\frac{d
\hat{V}}{dt}=\hat{F}(t)-\int_{0}^tdt'[\hat{F}(t),H_{ep}(t')],\label{r}
\end{eqnarray}
where $H_{ep}$ is the last term of (\ref{hami}), and $\hat{F}(t)$ is
the fluctuating force operator defined as
\begin{eqnarray}
\hat{F}(t)=-ig\sum_{q'}e^{i{q'}{\hat{R}}}{q'}u_{q'}\tilde{\rho}_{q'}.
\end{eqnarray}
Since $q$ is small, we can expand the exponential term to the linear
order, that is, $\exp(iq{\hat{R}})=1+iq{\hat{R}}$. Then Eq.
(\ref{r}) becomes
\begin{eqnarray}
\frac{d
\hat{V}}{dt}=\hat{F}(t)-\int_{0}^tdt'g^2(\hat{R}(t)+\hat{R}(t'))\sum_qq^2u^2_q(t,t')\cdot[\tilde{\rho}_q(t),\tilde{\rho}_q(t')],
\end{eqnarray}
where the zero-order term is absorbed into the fluctuating force,
and the high-order term of $q$ is neglected. $q'$ is equal to $q$ in
order to ensure the conservation of the total momentum.

Following the Ehrenfest theorem, the dynamics of classical phonons
could be treated as $u_q^2(t,t')\simeq k_BT\sin(\omega (t-t'))/K$
with $\omega$ its frequency and $K$ the elastic constant.\cite{JCP}
The last term, say $[\tilde{\rho}_q(t),\tilde{\rho}_q(t')]$, behaves
as a relaxation term and is significant when $t=t'$. It is not easy
to derive an explicit expression for this term, but it could be
estimated to be $\sim n/q^2$ with $n$ the density of
electron.\cite{GLE} In all, we then define a dimensionless parameter
$\xi$, which denotes the thermal average of the summation,
i.e.,$\langle
\sum_qq^2u^2_q(t,t')\cdot[\tilde{\rho}_q(t),\tilde{\rho}_q(t')]\rangle=\xi\sin(\omega(t-t'))$.
Actually, $\xi$ equals to the intensity of the fluctuating force and
is proportional to $k_BT$.\cite{GLE,JCP} Finally, by taking the
thermal average and using integration by parts, Eq. (\ref{r})
becomes the GLE as
\begin{eqnarray}
\frac{d \hat{V}}{dt}=\hat{F}(t)-\xi
g^2[2\hat{R}(t)-\int_{0}^tdt'\cos(\omega(t-t'))\hat{V}(t')],
\end{eqnarray}
where we have used $\hat{V}(t)=d\hat{R}(t)/dt$, and the cosine term
is absorbed into the fluctuating force. Here $\xi$ is an adjustable
parameter, and when it is sufficiently small, $\hat{R}(t)\simeq
\hat{V}(t)t$. To compare with the numerical results, we will set
$\xi$ to be 0.04. Then we take the thermal average for all the
quantities and make $\langle \hat{F}\rangle$ vanishing. The
numerical integration is now available for $V(t)$ when we consider
$V(0)=1$ as the initial condition.